\documentclass[preprint,aps,onecolumn,amsmath,amssymb,floatfix,endfloats,final]{revtex4}
\usepackage{graphicx}
\usepackage{dcolumn}
\usepackage{bm}

\begin{document}

\title{Electrically probing photonic bandgap phenomena in contacted defect nanocavities}

\author{F. Hofbauer}\email{hofbauer@wsi.tum.de}
\author{S. Grimminger}
\author{J. Angele}
\author{G. B\"ohm}
\author{R. Meyer}
\author{M.C. Amann}
\author{J. J. Finley}
\affiliation{Walter Schottky Institut and Physik Department, Technische Universit\"at M\"unchen,\\
Am Coulombwall 3, D-85748 Garching, Germany}

\date{\today}

\begin{abstract}
We demonstrate an electrically tunable two dimensional photonic crystal nanocavity containing InAs self assembled quantum dots. Photoluminescence and electroluminescence measurements are combined to probe the cavity mode structure and demonstrate a local electrical contact to the quantum dots. Measurements performed as a function of the electric field enable us to probe the capture, relaxation and recombination dynamics of photogenerated carriers inside the quantum dots emitting into a modified photonic environment. Furthermore, the two dimensional photonic crystal is probed by spatially dependent photocurrent spectroscopy indicating a $3.5\times$ enhancement of the local radiative lifetime of the QDs inside the photonic crystal environment.
\end{abstract}

\maketitle

Nanoengineered photonic materials currently attract widespread interest due to their ability to modify and control spontaneous emission and spatially redirect light from solids.\cite{Kaniber2007} Electrically contacted systems are required for applications in optoelectronics, such as high efficiency LEDs\cite{Oder2004} and low threshold nanolasers.\cite{Ellis2007,Hofmann2007,Strauf2006a}  However, the need for an electrical contact is less apparent for solid state quantum optics experiments involving quantum dot (QD) emitters\cite{Chang2006,Yoshie2004,Reithmaier2004,Hennessy2007}, where control of the spectral detuning between the QD emitter and the cavity mode ($\Delta\omega=\omega_{QD}-\omega_{cav}$) is needed. To date, a number of ingenious methods have already been developed to coarse tune $\omega_{cav}$ in two dimensional photonic crystal (PhC) defect nanocavities.\cite{Hennessy2005,Strauf2006b} Subsequently, $\Delta\omega$ can be \textit{fine tuned} by varying the temperature to shift $\omega_{QD}$ into resonance with $\omega_{cav}$. One drawback of temperature tuning is that large shifts are needed ($\Delta T\sim30-50$~K) to tune $\hbar\Delta\omega$ by a few meV.\cite{Yoshie2004,Reithmaier2004,Hennessy2007} Since the exciton decoherence rate increases rapidly with temperature\cite{Borri2005} methods are required to fine tune $\Delta\omega$ whilst preserving coherence.  Electrically contacted cavity-QD systems would enable reversible fine tuning of $\omega_{QD}$ using the quantum confined Stark effect\cite{Fry2001} and provide a route toward more sophisticated experiments.
\\
In this letter we demonstrate the fabrication and optical investigation of electrically contacted PhC defect nanocavities containing an ensemble of InAs self-assembled QDs.  We show that electrical contacts can be established by introducing doped contacts into the waveguide core to form $p$-$i$-$n$ nanocavity photodiodes. Emission and absorption spectroscopy allow us to probe the competition between carrier capture, relaxation, spontaneous emission and tunneling escape dynamics in a tailored photonic environment.

\begin{figure}[ht]
\centering
\includegraphics{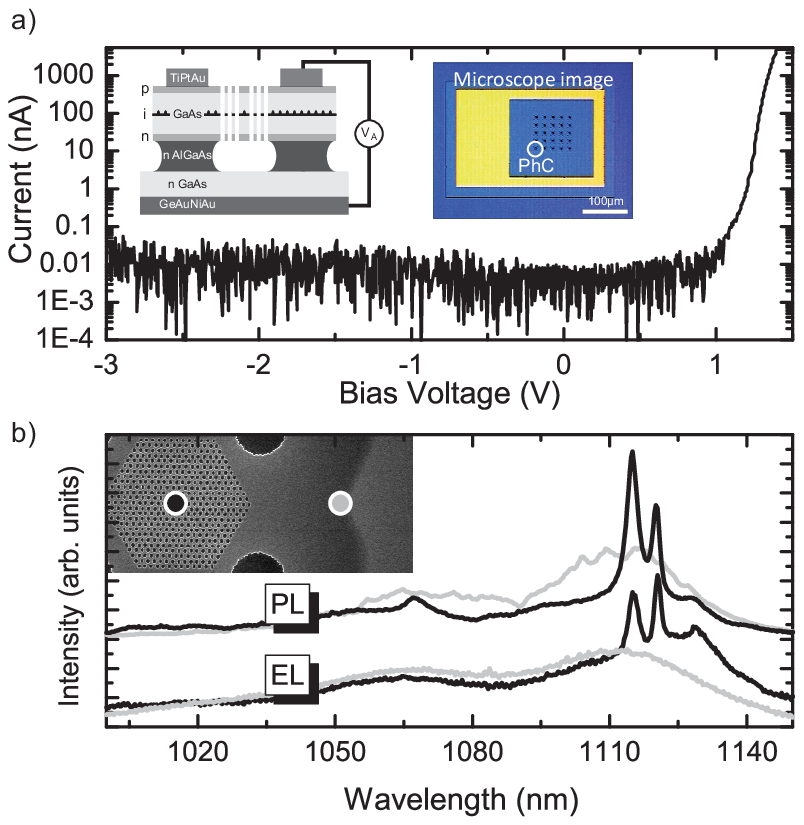}
\caption{\label{fig1}(a) Typical $I$-$V$ characteristics recorded from the $p$-$i$-$n$ nanocavity photodiodes, (inset) Schematic device structure and microscope image of processed samples. (b) Comparison of $\mu~PL$ and $\mu~EL$ measurements obtained \textit{on} the nanocavity and \textit{off} the PhC on the unpatterned membrane}
\end{figure}

The samples studied were $\mathrm{(Al)GaAs}$ \textit{p-i-n} photodiode structures and nominally consisted of the following layers: Firstly, we deposited a 750~nm thick, n-type ($n=2\times10^{18}cm^{-3}$, Si-dopant) $\mathrm{Al_{0.8}Ga_{0.2}As}$ sacrificial layer followed by a 30~nm thick, n-type GaAs lower contact layer ($n=2\times10^{18}cm^{-3}$). This was followed by the 110~nm thick intrinsic GaAs waveguide core into the center of which we grew a single layer of QDs by depositing 2.7 ML of InAs at $540^{\circ}C$ and a rate of $0.03$~ML/s.  A 30~nm thick p-type ($p=2\times10^{18}cm^{-3}$, Be-dopant) GaAs top contact was then grown to complete the structure.  After growth we established an Ohmic back contact to the buried $n^+$ layers and defined 250$\times$400 $\mu$m photodiode mesas using photolithography and wet etching techniques.  200$\times$200$\mu$m windows were opened in the top metallic Ohmic contact for optical access. Inside each window we then formed an array of 25 PhC nanocavities by combining electron beam lithography and $\mathrm{Cl_2}-\mathrm{Ar}$ reactive ion etching to define a lattice of cylindrical air holes (radius $r$) arranged in a hexagonal lattice with period $a = 280$~nm and $r/a = 0.3$. \cite{Kress2005b} Nanocavities were formed by single missing hole defects with reduced symmetry according to ref. \cite{Vuckovic2004}.  Finally, suspended PhC membrane structures were fabricated by selectively removing the $\mathrm{Al_{0.8}Ga_{0.2}As}$ layer beneath the GaAs waveguide core to leave a free standing, $p$-$i$-$n$ doped GaAs membrane.
\\
Figure \ref{fig1}(a) insert shows a schematic representation and optical microscope image of the nanocavity photodiode structures. Such devices enable tuning of the static electric field applied along the QD growth direction by tuning the applied bias ($V_A$) across the $p$-$i$-$n$ junction. A typical $I-V$ characteristic is presented in figure \ref{fig1}(a), revealing clear diode like behavior as expected with low leakage currents ($< 100$~pA) in the reverse bias direction and an onset of current flow close to $V_B\sim1.1\pm0.2$~V in the forward bias direction.  The relationship between $V_A$ and the static electric field can be approximated by $F = (V_B - V_A)/w$, where $w$ is the width of the intrinsic region.
\\
All optical experiments were performed at 10~K in a confocal microscopy setup that provides $\sim 1\mu$m spatial resolution. Figure \ref{fig1}(b) compares the inhomogeneously broadened QD photoluminescence ($PL$) spectra recorded for detection positions both \textit{on} the nanocavity and \textit{off} the PhC on the unpatterned membrane.  Both measurements were obtained with \textbf{non-resonant} optical excitation (633~nm, $\sim$~200~W/cm$^{-2}$). In both measurements emission from $s$-, $p$- and $d$- shells of the QDs are observed at $\sim 1120$~nm, $\sim 1060$~nm and $\sim 1010$~nm, respectively. When detected on the nanocavity the dipole like cavity modes are observed.\cite{Vuckovic2004} 
\\
In order to check that a local electrical contact is established to the nanocavity we turned off the laser, forward biased the $p$-$i$-$n$ diode and recorded electroluminescence ($EL$) spectra with a drive current of $I_{EL} = 35$~mA (see fig \ref{fig1}(b)). The $EL$ spectrum exhibits the same characteristic cavity modes which are absent from the measurement recorded away from the cavity.  These observations confirm that charge carriers can be injected into the nanocavity through the perforated PhC membrane and that an electrical contact is established to the dots within the structure. 

\begin{figure}[ht]
\centering
\includegraphics{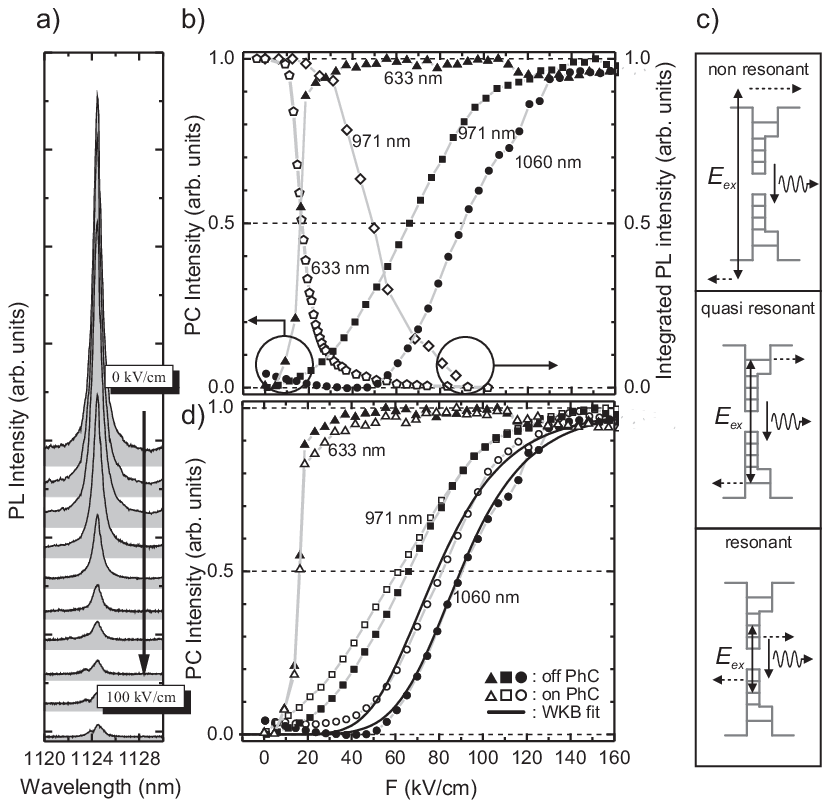}
\caption{\label{fig2}(a) $PL$ measurements recorded from a cavity mode as a function of the electric field; (b) Electric field dependencies of the $PC$ and $PL$ intensities for non-resonant (633~nm), quasi-resonant (971~nm) and resonant (1060~nm) excitation. (c) Schematic representations of the excitation schemes for non-resonant, quasi-resonant and p-shell excitation (d) Comparison of $PC$ measurements recorded on the PhC and unpatterned membrane revealing the shift of $F_{1/2}$ due to cavity QED effects.}
\end{figure}

We then \textit{reverse} biased the sample and performed complementary $PL$ and photocurrent ($PC$) measurements as a function of the electric field in the range $0$~V/cm to $100$~kV/cm.  Three distinct excitation regimes were explored; \textbf{non-resonant} excitation at 633~nm above the bandgap of the PhC matrix material, \textbf{quasi-resonant} at 971~nm into the excited states of the QDs and continuum states\cite{Vasinelli2001} and \textbf{resonant} into the p-shell of the QDs at 1060~nm.  Figure \ref{fig2}(a) shows typical field dependent $PL$ spectra recorded from one of the cavity modes under \textbf{non-resonant} excitation. Clearly, the $PL$ is strongest close to the flat band condition ($F = 0$~kV/cm, $V_A = 1.1$~V) and decreases rapidly with increasing field. The quantitative dependence of the integrated $PL$ intensity is plotted in figure \ref{fig2}(b), revealing this rapid quenching behavior.  Our findings for non-resonant excitation are compared with similar measurements recorded with \textbf{quasi-resonant} excitation. Both curves exhibit the same qualitative behavior, but higher electric fields are required to suppress the $PL$ for quasi-resonant excitation.  We quantified these observations by measuring the electric field where the $PL$ reaches half its initial intensity denoted by $F_{1/2}$.  We measure $F_{1/2}^{non-resonant}=16.9\pm0.2$~kV/cm and $F_{1/2}^{quasi-resonant}=49.1\pm0.2$~kV/cm, respectively. The observed value of $F_{1/2}^{non-resonant}<F_{1/2}^{quasi-resonant}$ since the intensity of the $PL$ reflects a competition between the various timescales for carrier capture, relaxation and recombination, and that for tunneling escape.  Clearly, the observed shift of $F_{1/2}$ to higher field for \textbf{quasi-resonant} excitation, when compared with \textbf{resonant} excitation, arises since carrier capture is suppressed by the internal electric field. For \textbf{resonant} excitation, the $PL$ pathway does not include capture (see figure \ref{fig2}(c)) the $PL$ intensity more closely reflects a trade off between carrier spontaneous recombination and tunneling escape.\cite{Fry2000}
\\
In order to cleanly probe the competition between spontaneous emission and carrier tunneling escape we performed $PC$ measurements to complement the $PL$ investigations. 
The field dependent $PC$ measured with \textbf{non-resonant}, \textbf{quasi-resonant} and \textbf{resonant} excitation is plotted as a function of electric field in \ref{fig2}(b). For each of these excitation regimes the $PC$ increases with $F$ as the system is driven from the situation where carriers are inefficiently extracted into the contacts ($F=0$~kV/cm) to that where \textit{all} the photogenerated carriers tunnel out of the dots and the $PC$ saturates ($F\sim60-150$~kV/cm).  Maintaining the notation $F_{1/2}^{PC}$ to represent the field where the \textit{PC} reaches 50\% of its saturation value, $F_{1/2}^{PC}$ shifts to higher electric field as the excitation becomes more resonant, with values of $16\pm0.5$~kV/cm, $66\pm 0.5$~kV/cm and $91\pm 0.5$~kV/cm having been measured for the three excitation regimes, respectively. The increase of $F_{1/2}^{PC}$ for tunneling out of the QDs arises since carriers excited more deeply within the dots tunnel less efficiently.  
For \textbf{non-resonant} excitation the normalized $PL$ and $PC$ curves intersect at a value of $\sim 0.5$, indicating that all charge carriers which do not become captured into the dots and recombine optically are extracted into the contacts.  In contrast, for quasi-resonant excitation the $PL$ and $PC$ intensity traces intersect at $\sim0.35$ indicating the presence of a more complicated relaxation pathway via multiple levels.  
\\
For resonant excitation the effective tunneling time primarily competes with radiative recombination, as discussed above, and the value of $F_{1/2}^{PC}$ provides an \textit{electrical measure} for the radiative lifetime of the QDs. In figure \ref{fig2}(d) we compare $PC$ measurements recorded on the unpatterned membrane with similar curves recorded from the PhC. The qualitative field dependence of both measurements are identical but $F_{1/2}^{PC}$ clearly shifts to lower fields by $\Delta F=4.0\pm 0.5$~kV/cm and $10.1\pm 0.5$~kV/cm for \textbf{quasi resonant} and \textbf{resonant} excitation, respectively. The observation of a shift of $F_{1/2}^{PC}$ to lower field indicates that the spontaneous emission time becomes longer on the photonic crystal compared with the unpatterned membrane. We identify this shift as arising from an enhanced radiative lifetime of QDs emitting inside the 2D photonic bandgap.\cite{Kaniber2007}  
For \textbf{quasi-resonant} excitation the tunneling competes with carrier relaxation process as discussed above, and $F_{1/2}^{PC}$ shifts only by $\sim4$~kV/cm. In complete contrast, for non-resonant excitation no shift of $F_{1/2}^{PC}$ is observed since tunneling escape competes with carrier capture, a process that does not depend on the local density of photonic states and is, therefore, the same, irrespective of the local photonic environment.
\\
To support these conclusions we modeled the observed $PC$ behavior using an adiabatic approximation with decoupled $z$ (parallel to the field) and $x$, $y$ (in the growth plane) motion. With this approximation, the tunneling probability depends only on the $z$ component of the wave function, and can be modeled using the one-dimensional (1D) WKB approximation.  For a 1D confining potential of width $L$ in a perpendicular field $F$, the tunneling rate $R_T$ is given by

\begin{equation}
R_T = \frac{\hbar\pi}{2m^* L^2}  \exp\left[ \frac{-4}{3\hbar e F} \sqrt{(2m^* E_I^3 )}\right],
\end{equation}

where $E_I$ is the ionization energy of the electron eigenstate.\cite{Fry2001} The normalized $PC$ signal ($I_{PC}$) can then be written as

\begin{equation}
\frac{I_{PC}}{I_{sat}}=\frac{\tau_{rec}}{\tau_{rec} + \tau_{esc}} = \frac{\tau_{rec} R_T}{1+\tau_{rec} R_T}
\end{equation}

to reflect the competition between radiative recombination and the local spontaneous emission lifetime ($\tau_{rec}$).  We fitted this relation to the $PC$ curves recorded from the unpatterned membrane using a measured intrinsic spontaneous recombination time of $\tau_{rec} = 1$~ns\cite{Kaniber2007} and adjusted the dot height, ionization energy and effective mass to obtain a best fit ($L=35\pm 5$~{\AA}, $E_I=190 \pm 30$~meV and $m^*=0.049\pm 0.005 m^0$).  The same parameters were then used to fit the data recorded from the PhC, changing only the parameter $\tau_{rec}$ to obtain the two fits presented on figure \ref{fig2}(d). Using this method, we obtain $\tau_{on}/ \tau_{off} = 3.5\pm 0.3$ demonstrating that the photonic bandgap significantly suppresses spontaneous emission, in good agreement with recent time resolved spectroscopy measurements on similar structures.\cite{Kaniber2007}
\begin{figure}[ht]
\centering
\includegraphics{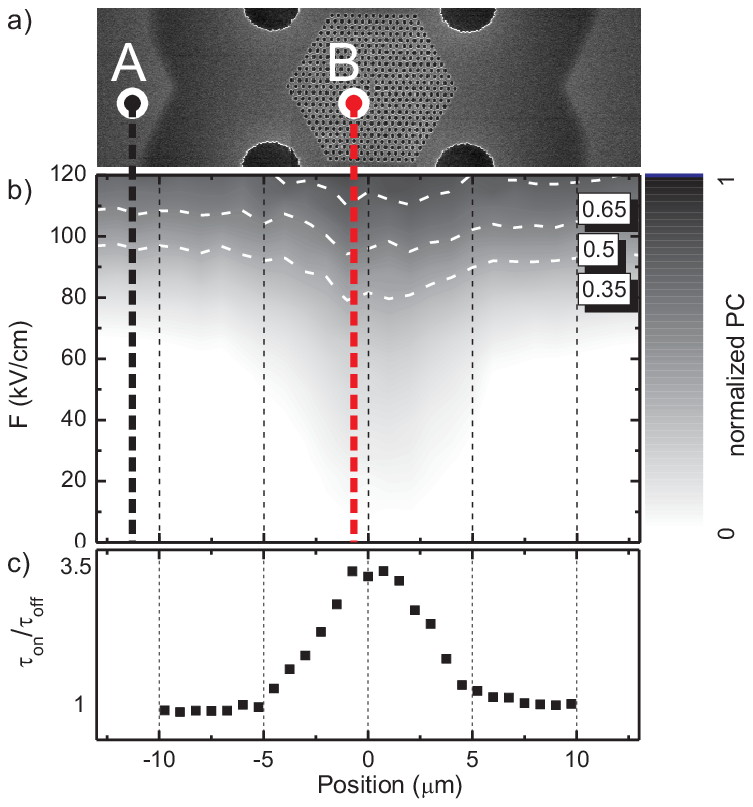}
\caption{\label{fig3}(a) (b) Spatial mapping of the $PC$ for different electric fields. The points labeled \textbf{A} and \textbf{B} correspond to the detection positions presented in figure \ref{fig2}(d), on and off the PhC. (c) Local radiative lifetime extracted from the WKB analysis discussed in the text.}
\end{figure}
To further confirm that the observed phenomena arises from the 2D photonic bandgap, we spatially mapped across the the $PC$ region and recorded the field dependent $PC$ curves when subject to resonant excitation.  The results of these measurements are presented in fig. \ref{fig3}(a). as a the false color image of $I_{PC}(F)$ data versus detection position. Outside the PhC the $I_{PC}(F)$ curves are independent of detection position since the radiative lifetime is $\sim 1$~ns. In contrast, when exciting on the PhC the $I_{PC}(F)$ traces systematically shift to lower field, the strongest shift being observed for $F\sim10$~kV/cm in the center of the structure. Fitting the data with the WKB method discussed above allows us to extract the local spontaneous emission lifetime.  The results of this analysis are presented in figure \ref{fig3}(c), revealing a systematic lengthening of the spontaneous emission lifetime when detecting upon the photonic crystal.  
\\
In summary, we have realized electrically contacted two-dimensional PhC defect nanocavities containing an ensemble of InAs self-assembled QDs and studied their fundamental optical properties.  
The 2D photonic bandgap is probed using a fully electrical approach and comparison of our results with calculations indicate that the spontaneous emission lifetime is suppressed by a factor $\sim3.5\times$ by the 2D photonic bandgap.
\\
We acknowledge financial support of the Deutsche Forschungsgemeindschaft via the Sonderforschungsbereich 631, Teilprojekt B3 and the German Excellence Initiative via the "Nanosystems Initiative Munich (NIM)".

\end{document}